\documentclass[a4paper,12pt]{article}
\usepackage{mathptmx}       
\usepackage{helvet}         
\usepackage{courier}        
\usepackage{type1cm}        
 \usepackage{wrapfig}                           
\usepackage{latexsym}
\usepackage{amsfonts}
\usepackage{makeidx}         
\usepackage{graphicx}        
\usepackage{multicol}        
\usepackage[bottom]{footmisc}%

\title{Integrable Nonlocal Reductions}

\author{Metin G\"{u}rses \thanks{gurses@fen.bilkent.edu.tr}\\
{\small Department of Mathematics, Faculty of Science}\\
{\small Bilkent University, 06800 Ankara - Turkey}\\
Asl{\i} Pekcan \thanks{Email:aslipekcan@hacettepe.edu.tr} \\
{\small Department of Mathematics, Faculty of Science} \\
{\small Hacettepe University, 06800 Ankara - Turkey}
}

\date{\nonumber}
\begin{document}
\maketitle
\date{\nonumber}
\newtheorem{thm}{Theorem}[section]
\newtheorem{Le}{Lemma}[section]
\newtheorem{defi}{Definition}[section]
\newtheorem{ex}{Example}[section]
\newtheorem{pro}{Proposition}[section]
\baselineskip 17pt

\begin{abstract}
 We present some nonlocal integrable systems  by using the Ablowitz-Musslimani nonlocal reductions.
We first present all possible nonlocal reductions of nonlinear Schr\"{o}dinger (NLS) and modified Korteweg-de Vries (mKdV) systems. We give soliton solutions of these nonlocal equations by using the Hirota method. We extend the nonlocal NLS equation to nonlocal Fordy-Kulish equations by utilizing the nonlocal reduction to the Fordy-Kulish system on symmetric spaces.  We also consider the super AKNS system and then show that Ablowitz-Musslimani nonlocal reduction can be extended to super integrable equations.
We obtain new nonlocal equations namely nonlocal super NLS and nonlocal super mKdV equations.
\end{abstract}

\section{Introduction}

After the publications of the Ablowitz-Musslimani works \cite{AbMu1}-\cite{AbMu3} on nonlocal nonlinear Schr\"{o}dinger (NLS) equation
there is a huge interest in obtaining  nonlocal reductions of systems of integrable equations \cite{fok}-\cite{chen}. In all these works the soliton solutions and their properties were investigated by using inverse scattering techniques, by Darboux transformations, and by the Hirota direct method.

\vspace{0.3cm}

\noindent Recently we extended the nonlocal NLS equations to nonlocal Fordy-Kulish equations by utilizing the nonlocal reduction to the Fordy-Kulish system on symmetric spaces \cite{GursesFK}. In a previous work \cite{GurPek1} we studied the coupled NLS system obtained from AKNS scheme. By using the Hirota bilinear method we first found soliton solutions of the coupled NLS system of equations then using the Ablowitz-Musslimani type reduction formulas we obtained the soliton solutions of the standard and time T-, space S-, and space-time ST- reversal symmetric nonlocal NLS equations. Similarly, in a recent work \cite{GurPek2} we studied the nonlocal modified Korteweg-de Vries (mKdV) equations which are also obtained from
AKNS scheme by Ablowitz-Musslimani type nonlocal reductions. For this purpose we start using the soliton solutions  of the
coupled mKdV system found by Hirota and Iwao \cite{IWHI}. Then by using these solutions and Ablowitz-Musslimani type reduction formulas we obtained solutions of standard and nonlocal mKdV and complex mKdV (cmKdV) equations including one-, two-, and three-soliton waves, complexitons, breather-type, and kink-type waves. We used two different types of approaches in finding the soliton solutions. We gave one-soliton solutions of both types and presented only first type of two- and three-soliton solutions (see \cite{GurPek2}).\\

\noindent When the Lax pair, in $(1+1)$-dimensions, is given in a Lie algebra the resulting evolution equations are given as
a coupled system
\begin{eqnarray}
q^{i}_{t}&=&F^{i}(q^{k}, r^{k}, q^{k}_{x}, r^{k}_{x}, q^{k}_{xx}, r^{k}_{xx}, \cdots),~~~  \label{cdenk1}\\
 \nonumber \\
r^{i}_{t}&=&G^{i}(q^{k}, r^{k}, q^{k}_{x}, r^{k}_{x}, q^{k}_{xx}, r^{k}_{xx}, \cdots), \label{cdenk2}
\end{eqnarray}
for all $i=1,2,\cdots,N$ where $F^{i}$ and $G^{i} ~(i=1,2,\cdots,N)$ are functions of the dynamical variables $q^{i}(t,x)$, $r^{i}(t,x)$, and their partial derivatives with respect to $x$. Since we start with a Lax pair then the system (\ref{cdenk1})-(\ref{cdenk2}) is an integrable system of nonlinear partial differential equations.\\

\noindent In the space of dynamical variables $(q^{i},r^{i})$ there exist subspaces
\begin{equation}
r^{i}(t,x)=k q^{i}(t,x),
\end{equation}
or
\begin{equation}
r^{i}(t,x)=k \bar{q}^{i}(t,x),
\end{equation}
where $k$ is a constant and a bar over a letter denotes complex conjugation, such that the systems of equations (\ref{cdenk1})-(\ref{cdenk2}) reduce to one system for $q^{i}$'s
\begin{equation}\label{first}
q^{i}_{t}={\tilde F}^{i}(q^{k}, q^{k}_{x}, q^{k}_{xx}, \cdots)
\end{equation}
provided that the second system (\ref{cdenk2}) consistently reduces to the above system (\ref{first}) of equations. Here ${\tilde F}=F|_{r=k \bar{q}}$.
Recently a new reduction is introduced by Ablowitz and Musslimani \cite{AbMu1}-\cite{AbMu3}
\begin{equation}
r^{i}(t,x)=k q^{i}(\mu_{1} t, \mu_{2} x), \label{red1}
\end{equation}
or
\begin{equation}
r^{i}(t,x)=k \bar{q}^{i}(\mu_{1} t, \mu_{2} x), \label{red2}
\end{equation}
for $i=1,2,\cdots,N$. Here $k$ is a constant and $\mu_{1}^2=\mu_{2}^2=1$. When
$(\mu_{1}, \mu_{2})=\{(-1,1),\\(1,-1),(-1,-1)\}$ the above constraints reduce the system (\ref{cdenk1}) to nonlocal differential equations provided that the second system (\ref{cdenk2}) consistently reduces to the first one. If the reduction is done in a consistent way the reduced system of equations is also integrable. This means that the reduced system
admits a recursion operator and bi-hamiltonian structure and the reduced system has $N$-soliton solutions. The inverse scattering method (ISM) can also be applied. Ablowitz and Musslimani have first found the nonlocal NLS equation from the coupled AKNS equations and solved it by ISM \cite{AbMu2}.

In our studies of nonlocal NLS and nonlocal mKdV equations we introduced a general method to obtain soliton solutions of nonlocal integrable equation. This method consists of  three main steps:,

\vspace{0.3cm}
\noindent
{\bf (i)} Find a consistent reduction formula which reduces the integrable system of equations to integrable nonlocal equations.

\vspace{0.3cm}
\noindent
{\bf (ii)} Find soliton solutions of the system of equations by use of the Hirota direct method or by inverse scattering transform technique, or by use of Darboux Transformation.

\vspace{0.3cm}
\noindent
{\bf (iii)} Use the reduction formulas on the soliton solutions of the system of equations to obtain the soliton solutions of the reduced nonlocal equations. By this way one obtains many different relations among the soliton parameters of the system of equations.

\vspace{0.3cm}
\noindent
In the following sections we mainly follow the above method in obtaining the soliton solutions of the nonlocal NLS and nonlocal mKdV equations.

\section{AKNS System}

When we begin with the Lax pair in $sl(2,R)$ algebra and assume them as a polynomial of the spectral parameter of degree less or equal to three then we obtain the following system of evolution equations \cite{AKNS}:
\begin{eqnarray}
&& q_{t}=a_{2}\, (-\frac{1}{2}\, q_{xx}+q ^2\, r)+i a_{3}\,(-\frac{1}{4}\, q_{xxx} +\frac{3}{2}\, q r q_{x}), \label{eq1} \\
&& r_{t}=a_{2}\, (\frac{1}{2}\, r_{xx}-q \, r^2)+i a_{3}\,(-\frac{1}{4}\, r_{xxx}+\frac{3}{2}\, q r r_{x}). \label{eq2}
\end{eqnarray}
Here $a_{2}$ and $a_{3}$ are arbitrary constants.\\

\noindent Letting $a_{2}=1/a$ and $a_{3}=0$ we get the coupled NLS system,
\begin{eqnarray}
&& a\,q_{t}=-\frac{1}{2}\, q_{xx}+q ^2\, r, \label{eq7} \\
&& a\, r_{t}=\frac{1}{2}\, r_{xx}-q \, r^2, \label{eq8}
\end{eqnarray}
where $a$ is any constant. The corresponding recursion operator is
\begin{equation}\label{recur}
{\cal R}=\left(\begin{array}{ll} q\,D_{x}^{-1}\,r-\frac{a}{2}\, D_{x} &\quad\quad q\, D_{x}^{-1}\,q \cr
-r\,D_{x}^{-1}\, r &\quad -r\, D_{x}^{-1}\,q+\frac{a}{2}\,D_{x}
\end{array}\right).
\end{equation}
One-soliton solution of the system (\ref{eq7})-(\ref{eq8}) can be obtained by the Hirota method as
\begin{equation}\label{NLSsystemonesol}
\displaystyle q(t,x)=\frac{e^{\theta_1}}{1+Ae^{\theta_1+\theta_2}}, \quad \quad r(t,x)=\frac{e^{\theta_2}}{1+Ae^{\theta_1+\theta_2}},
\end{equation}
where $\theta_{i}=k_{i} x+ \omega_{i} t +\delta_{i}$, $i=1,2$ with $\omega_{1}=k_{1}^2/2a$, $\omega_{2}=-k_{2}^2/2a$, and $A=-1/(k_{1}+k_{2})^2$. Here $k_{1}$, $k_{2}$, $\delta_{1}$, and $\delta_{2}$ are arbitrary complex numbers.

\section{Standard and Nonlocal NLS Equations}
Standard reduction of NLS equation is $r(t,x)=k \bar{q}(t,x)$ where $k$ is a real constant. The second equation (\ref{eq8}) is consistent
if $\bar{a}=-a$. Then the NLS system reduces to
\begin{equation}
a\,q_{t}=-\frac{1}{2}\, q_{xx}+k\,q ^2\, \bar{q}. \label{eq5}
\end{equation}

\noindent Recursion operator of the NLS equation is
\begin{equation}
{\cal R}=\left(\begin{array}{ll} q\,D_{x}^{-1}\,\bar{q}-\frac{a}{2}\, D_{x} &\quad\quad q\, D_{x}^{-1}\,q \cr
-\bar{q}\,D_{x}^{-1}\, \bar{q} &\quad -\bar{q}\, D_{x}^{-1}\,q+\frac{a}{2}\,D_{x}
\end{array}\right).
\end{equation}
There are two types of approaches to find solutions
of the standard and nonlocal NLS equations. In Type 1, one-soliton solution is obtained by letting $k_{2}=\bar{k}_{1}$ and $e^{\delta_{2}}=k e^{\bar{\delta}_{1}}$ in (\ref{NLSsystemonesol}) as
\begin{equation}
q(t,x)=\frac{e^{\theta_{1}}}{1+A\,k\, e^{\theta_{1}+\bar{\theta}_{1}}}.
\end{equation}
\vspace{0.2cm}
In Type 2 we obtain a different solution under the constraints,
\begin{equation}
1)\,\, \bar{a}=-a,\,\, 2)\,\, k_1=-\bar{k}_1,\,\, 3)\,\, k_2=-\bar{k}_2,\,\, 4)\,\, Ake^{\delta_1+\bar{\delta}_1}=1,\,\, 5)\,\, Ae^{\delta_2+\bar{\delta_2}}=k.
\end{equation}
If we take $a=i\alpha$, $k_1=i\beta$, $k_2=i\gamma$, $e^{\delta_1}=a_1+ib_1$, and $e^{\delta_2}=a_2+ib_2$ for $\alpha, \beta, \gamma, a_j, b_j$\\
$\in \mathbb{R}$, $j=1, 2$ one-soliton solution of standard NLS equation becomes
\begin{equation}\displaystyle
q(t,x)=\frac{e^{i\beta x+\frac{i\beta^2}{2\alpha}t}(a_1+ib_1)}{1+\frac{1}{(\beta+\gamma)^2}e^{i(\beta+\gamma)x+i\frac{(\beta^2-\gamma^2)}{2\alpha}t}(a_1+ib_1)(a_2+ib_2)},\quad \beta\neq-\gamma,
\end{equation}
and therefore
\begin{equation}\label{localONESOL}\displaystyle
|q(t,x)|^2=\frac{a_1^2+b_1^2}{4}\sec^2\Big(\frac{\theta}{2}\Big),
\end{equation}
where
$$ \theta=(\beta+\gamma)x+\frac{1}{2\alpha}(\beta^2-\gamma^2)t+\omega_0$$
for $\omega_0=\arccos((a_1a_2-b_1b_2)/(\beta+\gamma)^2)$ with $a_1^2+b_1^2=(\beta+\gamma)^2/k$ and $a_2^2+b_2^2=k(\beta+\gamma)^2$. This solution is singular for any choice of the parameters.\\

\noindent Let now $r(t,x)=k\, \bar{q}(\mu_{1} t, \mu_{2} x)$ where $\mu_{1}^2=\mu_{2}^2=1$
and $k$ is a real constant. This is an integrable reduction, meaning that the new equation we obtain
\begin{equation}
a\,q_{t}(t,x)=-\frac{1}{2}\, q_{xx}(t,x)+k\,q ^2(t,x)\, \bar{q}(\mu_{1} t, \mu_{2} x), \label{eq9}
\end{equation}
is integrable and the second equation (\ref{eq8}) is consistent with the first one (\ref{eq7}) provided that
$\bar{a}=-\mu_{1}\,a $. The recursion operator of this equation is
\begin{equation}
{\cal R}=\left(\begin{array}{ll} k\,q(t,x)\,D_{x}^{-1}\,\bar{q}(\mu_{1} t, \mu_{2} x)-\frac{a}{2}\, D_{x} &\quad\quad q(t,x)\, D_{x}^{-1}\,q(t,x) \cr
-k^2\,\bar{q}(\mu_{1} t, \mu_{2} x)\,D_{x}^{-1}\, \bar{q}(\mu_{1} t, \mu_{2} x) &\quad - k\,\bar{q}(\mu_{1} t, \mu_{2} x)\, D_{x}^{-1}\,q(t,x)+\frac{a}{2}\,D_{x}
\end{array}\right),
\end{equation}
and one-soliton solution is obtained by letting $k_{2}= \mu_{2}\,\bar{k}_{1}$ and $e^{\delta_{2}}=k e^{\bar{\delta}_{1}}$ in
(\ref{NLSsystemonesol}) as
\begin{equation}
q(t,x)=\frac{e^{\theta_{1}(t,x)}}{1+A\,k\, e^{\theta_{1}(t,x)+\bar{\theta}_{1}(\mu_{1}\,t, \mu_{2}\,x)}},
\end{equation}
in Type 1 approach.\\

\noindent In Type 2, under the constraints
\begin{equation}
1)\,\, \bar{a}=-\mu_{1}a, \,\, 2)\,\, \, k_1=-\bar{k}_1\mu_2, \,\, 3)\,\, k_2=-\bar{k}_2\mu_2,\,\, 4)\,\,  Ake^{\delta_1+\bar{\delta}_1}=1,
\,\, 5)\,\,  Ae^{\delta_2+\bar{\delta}_2}=k,
\end{equation}
we obtain a different one-soliton solution.\\

\noindent Nonlocal reductions of NLS system correspond to $(\mu_{1}, \mu_{2})=\{(-1,1), (1,-1), \\(-1,-1)\}$. Hence we have three different reductions of the NLS system (\ref{eq7})-(\ref{eq8}).\\

\noindent 1) T-Symmetric NLS Equations: Let $r(t,x)=k\, \bar{q}(-t,  x)$. This is an integrable equation
\begin{equation}
a\,q_{t}(t,x)=-\frac{1}{2}\, q_{xx}(t,x)+k\,q ^2(t,x)\, \bar{q}(-t,x), \label{eq9}
\end{equation}
 provided that $\bar{a}=a$. The recursion operator of this equation is
\begin{equation}
{\cal R}=\left(\begin{array}{ll} k\,q(t,x)\,D_{x}^{-1}\,\bar{q}(-t,x)-\frac{a}{2}\, D_{x} &\quad \quad q(t,x)\, D_{x}^{-1}\,q(t,x) \cr
-k^2\,\bar{q}(-t,x)\,D_{x}^{-1}\, \bar{q}(-t,x) &\quad - k\,\bar{q}(-t,x)\, D_{x}^{-1}\,q(t,x)+\frac{a}{2}\,D_{x}
\end{array}\right),
\end{equation}
and one-soliton solution is obtained by letting $k_{2}= \bar{k}_{1}$ where $k_1=\alpha+i\beta$, $\alpha, \beta \in \mathbb{R}$,
 and $e^{\delta_{2}}=k e^{\bar{\delta}_{1}}$ in (\ref{NLSsystemonesol}) as
\begin{equation}
\displaystyle q(t,x)=\frac{e^{(\alpha+i\beta)x+\frac{(\alpha+i\beta)^2}{2a}t+\delta_1}}{1-k\frac{e^{2\alpha x+\frac{2i\alpha\beta}{a}t+\delta_1+\bar{\delta}_1}}{4\alpha^2}},
\end{equation}
for $\alpha\neq 0$ in Type 1. To have a real-valued solution we consider $q(t,x)\bar{q}(t,x)=|q(t,x)|^2$.
Here we have
\begin{equation}\label{caseamoduleonesolnNLS}
\displaystyle |q(t,x)|^2=\frac{16\alpha^4e^{2\alpha x+\frac{\alpha^2-\beta^2}{a}t+\delta_1+\bar{\delta}_1}}{(ke^{2\alpha x+\delta_1+\bar{\delta}_1}
-4\alpha^2\cos(\frac{2\alpha\beta}{a}t) )^2+16\alpha^4\sin^2(\frac{2\alpha\beta}{a}t)}.
\end{equation}
When $\beta \ne 0$ and
$$\displaystyle t=\frac{a n \pi}{2 \alpha \beta},\quad ke^{2\alpha x+\delta_1+\bar{\delta}_1}-4\alpha^2\,(-1)^n=0,$$
where $n$ is an integer, both focusing (sign $(k)=-1$)  and defocusing (sign $(k)=1$) cases have singularities. When $\beta=0$ the focusing case is non-singular but asymptotically growing in time.\\

\noindent In Type 2, if we take $k_1=i\beta$, $k_2=i\gamma$ for $\beta,\gamma \in \mathbb{R}$, $e^{\delta_1}=a_1+ib_1$, and $e^{\delta_2}=a_2+ib_2$ for $a_j, b_j \in \mathbb{R}$, $j=1, 2$ then one-soliton solution becomes
\begin{equation}\displaystyle
q(t,x)=\frac{e^{i\beta x-\frac{\beta^2}{2a}t}(a_1+ib_1)}{1+\frac{1}{(\beta+\gamma)^2}e^{i(\beta+\gamma)x+\frac{(\gamma^2-\beta^2)}{2a}t}(a_1+ib_1)(a_2+ib_2)},\quad \beta\neq -\gamma.
\end{equation}
Hence the function $|q(t,x)|^2$ is
\begin{equation}\label{nonlocalONECASEA}\displaystyle
|q(t,x)|^2=
\frac{e^{\frac{(\gamma^2+\beta^2)}{2a}t}(a_1^2+b_1^2) }{2[\cosh(\frac{(\gamma^2-\beta^2)}{2a}t)+\cos\theta]},
\end{equation}
where
 $$\theta=(\beta+\gamma)x+\omega_0$$
for $\omega_0=\arccos((a_1a_2-b_1b_2)/(\beta+\gamma)^2)$ with $a_1^2+b_1^2=(\beta+\gamma)^2/k$ and $a_2^2+b_2^2=k(\beta+\gamma)^2$. Clearly, the solution is singular at $t=0$ and $\theta=(2n+1)\pi$, $n$ integer and
non-singular for $t\neq 0$.\\

\noindent 2) S-Symmetric NLS Equations: Let $r(t,x)=k\, \bar{q}(t,-x)$. This is an integrable equation
\begin{equation}
a\,q_{t}(t,x)=-\frac{1}{2}\, q_{xx}(t,x)+k\,q ^2(t,x)\, \bar{q}(t,-x), \label{eq9}
\end{equation}
 provided that $\bar{a}=-a$. The recursion operator of this equation is
\begin{equation}
{\cal R}=\left(\begin{array}{ll} k\,q(t,x)\,D_{x}^{-1}\,\bar{q}(t,-x)-\frac{a}{2}\, D_{x} &\quad \quad q(t,x)\, D_{x}^{-1}\,q(t,x) \cr
-k^2\,\bar{q}(t,-x)\,D_{x}^{-1}\, \bar{q}(t,-x) &\quad - k\,\bar{q}(t,-x)\, D_{x}^{-1}\,q(t,x)+\frac{a}{2}\,D_{x}
\end{array}\right).
\end{equation}
 In Type 1, one-soliton solution is obtained by letting $k_{2}= -\bar{k}_{1}$ where $k_1=\alpha+i\beta$, $\alpha, \beta \in \mathbb{R}$, $a=iy$, $y\in \mathbb{R}$, and $e^{\delta_{2}}=k e^{\bar{\delta}_{1}}$ in
(\ref{NLSsystemonesol}) as
\begin{equation}
\displaystyle q(t,x)=\frac{e^{(\alpha+i\beta)x+\frac{(\alpha+i\beta)^2}{2iy}t+\delta_1}}{1+k\frac{e^{2i\beta x+\frac{2\alpha\beta}{y}t+\delta_1+\bar{\delta}_1}}{4\beta^2}},
\end{equation}
where $\beta\neq 0$. Hence the function $|q(t,x)|^2$ is
\begin{equation}\label{casebmoduleonesolnNLS}
\displaystyle |q(t,x)|^2=\frac{16\beta^4e^{2\alpha x+\frac{2\alpha\beta}{y}t+\delta_1+\bar{\delta}_1}}{(ke^{\frac{2\alpha\beta}{y}t+\delta_1+\bar{\delta}_1} +4\beta^2\cos(2\beta x))^2+16\beta^4\sin^2(2\beta x)}.
\end{equation}
If $\alpha\neq 0$ the above function is singular at
$$\displaystyle x=\frac{n \pi}{2\beta},\quad ke^{\frac{2\alpha\beta}{y} t+\delta_1+\bar{\delta}_1} +4\beta^2\, (-1)^n=0,$$
where $n$ is an integer, both for focusing and defocusing cases. If $\alpha=0$, the function (\ref{casebmoduleonesolnNLS}) becomes
\begin{equation}\label{casebmoduleonesolnNLSalpha=0}\displaystyle
|q(t,x)|^2=\frac{2\beta^2}{k[B+\cos(2\beta x)]},
\end{equation}
for $ B=(\rho^2+16\beta^4)/(8\rho\beta^2)$ where $\rho=ke^{\delta_1+\bar{\delta}_1}$. Obviously, the solution (\ref{casebmoduleonesolnNLSalpha=0}) is non-singular if $B>1$ or $B<-1$.\\

\begin{wrapfigure}{r}{0.3\textwidth}
  \vspace{-21pt}
  \begin{center}
    \includegraphics[width=0.18\textwidth]{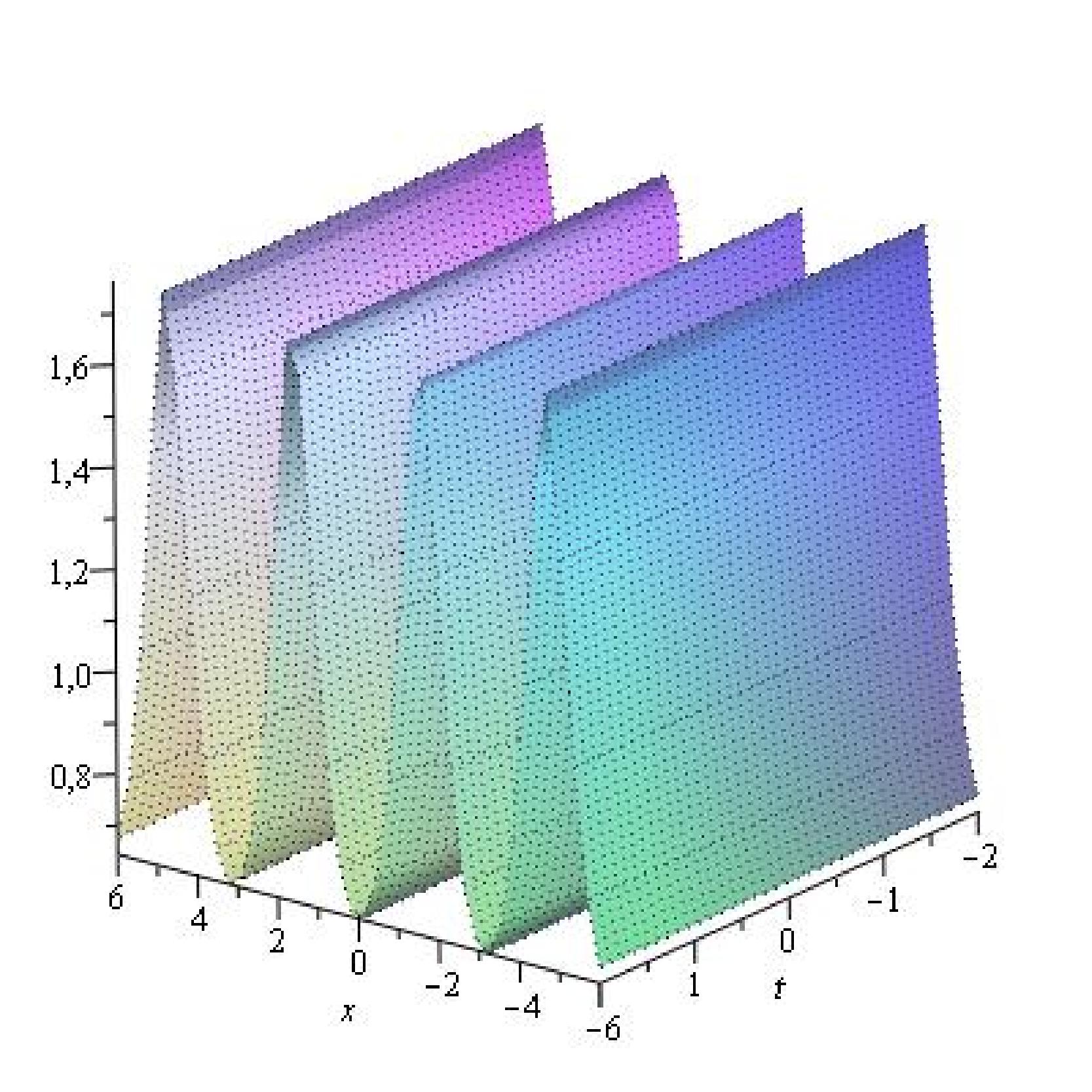}
  \end{center}
  \vspace{-18pt}
  \centering
  \caption{A periodic solution}
  \vspace{-10pt}
\end{wrapfigure}

\noindent \textbf{Example 1.} For the set of parameters $ (k_1, k_2, e^{\delta_1}, e^{\delta_2}, k, a)=(i, i, i, -i, 1, i/2)$, we get the solution
$$|q(t,x)|^2=\frac{16}{(17+8\cos(2x))}.$$ This solution represents a periodic solution. Its graph is given in Figure 1.\\

\noindent For Type 2 if we let $a=i\alpha$, $\alpha\in \mathbb{R}$, $e^{\delta_1}=a_1+ib_1$, and $e^{\delta_2}=a_2+ib_2$ for $a_j, b_j \in \mathbb{R}$, $j=1, 2$ then one-soliton solution becomes
\begin{equation}\displaystyle
q(t,x)=\frac{e^{k_1x+i\frac{k_1^2}{2\alpha}t}(a_1+ib_1)}{1-\frac{1}{(k_1+k_2)^2}e^{(k_1+k_2)x-i\frac{(k_1^2-k_2^2)}{2\alpha}t}(a_1+ib_1)(a_2+ib_2)},\quad k_1\neq -k_2.
\end{equation}
Therefore the function $|q(t,x)|^2$ is
\begin{equation}\label{nonlocalONECASEB}\displaystyle
|q(t,x)|^2=\frac{e^{(k_1-k_2)x}(a_1^2+b_1^2)}{2[\cosh((k_1+k_2)x)-\cos\theta]},
\end{equation}
where
$$ \theta=\frac{1}{2\alpha}(k_1^2-k_2^2)t-\omega_0$$
for $\omega_0=\arccos((a_1a_2-b_1b_2)/(k_1+k_2)^2)$ with $a_1^2+b_1^2=-(k_1+k_2)^2/k$ and $a_2^2+b_2^2=-k(k_1+k_2)^2$. The solution is singular at $x=0$ and $\theta=2n\pi$ for $n$ integer, and non-singular for
$x\neq 0$.\\

\noindent 3) ST-Symmetric  NLS Equations: Let $r(t,x)=k\, \bar{q}(-t,-x)$. This is an integrable equation
\begin{equation}
a\,q_{t}(t,x)=-\frac{1}{2}\, q_{xx}(t,x)+k\,q ^2(t,x)\, \bar{q}(-t,-x), \label{eq9}
\end{equation}
 provided that $\bar{a}=-a$. The recursion operator of this equation is
\begin{equation}
{\cal R}=\left(\begin{array}{ll} k\,q(t,x)\,D_{x}^{-1}\,\bar{q}(-t,-x)-\frac{a}{2}\, D_{x} &\quad \quad q(t,x)\, D_{x}^{-1}\,q(t,x) \cr
-k^2\,\bar{q}(-t,-x)\,D_{x}^{-1}\, \bar{q}(-t,-x) &\quad - k\,\bar{q}(-t,-x)\, D_{x}^{-1}\,q(t,x)+\frac{a}{2}\,D_{x}
\end{array}\right),
\end{equation}
and one-soliton solution is obtained by letting $k_{2}= -\bar{k}_{1}$ where $k_1=\alpha+i\beta$, $\alpha, \beta \in \mathbb{R}$ and $e^{\delta_{2}}=k e^{\bar{\delta}_{1}}$ in
(\ref{NLSsystemonesol}) as
\begin{equation}
\displaystyle q(t,x)=\frac{e^{(\alpha+i\beta)x+\frac{(\alpha+i\beta)^2}{2a}t+\delta_1}}{1+k\frac{e^{2i\beta x+\frac{2i\alpha\beta}{a}t+\delta_1+\bar{\delta}_1}}{4\beta^2}},
\end{equation}
where $\beta\neq 0$ in Type 1. Therefore $|q(t,x)|^2$ is
\begin{equation}\label{casecmoduleonesolnNLS}
\displaystyle |q(t,x)|^2=\frac{16\beta^4e^{2\alpha x+\frac{(\alpha^2-\beta^2)}{a}t+\delta_1+\bar{\delta}_1}}{(ke^{\delta_1+\bar{\delta}_1} +4\beta^2\cos(2\beta x+\frac{2\alpha\beta}{a}t))^2+16\beta^4\sin^2(2\beta x+\frac{2\alpha\beta}{a}t)}.
\end{equation}
This function is singular on the line $2\beta x+(2\alpha\beta t/ a)=n \pi$ where $n$ is an integer, if the condition
$ke^{\delta_1+\bar{\delta}_1} +4\beta^2\, (-1)^n=0$ is satisfied by the parameters of the solution, otherwise it represents a
non-singular wave solution for both focusing and defocusing cases. For $\alpha=0$, $(a>0)$, the solution represents a localized wave solution.\\

\noindent In Type 2, if we take $e^{\delta_1}=a_1+ib_1$ and $e^{\delta_2}=a_2+ib_2$ for $a_j, b_j \in \mathbb{R}$, $j=1, 2$ we have the one-soliton solution as
\begin{equation}\displaystyle
q(t,x)=\frac{e^{k_1x+\frac{k_1^2}{2a}t}(a_1+ib_1)}{1-\frac{1}{(k_1+k_2)^2}e^{(k_1+k_2)x+(\frac{(k_1^2-k_2^2)}{2a})t}(a_1+ib_1)(a_2+ib_2)},\quad k_1\neq -k_2.
\end{equation}
The corresponding function $|q(t,x)|^2$ is
\begin{equation}\label{nonlocalONECASEC}\displaystyle
|q(t,x)|^2=\frac{e^{\phi}(a_1^2+b_1^2)}{1-2\gamma e^{\theta}+e^{2\theta}},
\end{equation}
where
$$\phi=2k_1x+\frac{k_1^2}{a}t,\quad \theta=(k_1+k_2)x+\frac{1}{2a}(k_1^2-k_2^2)t,$$
$\gamma=(a_1a_2-b_1b_2)/(k_1+k_2)^2$, $a_1^2+b_1^2=-(k_1+k_2)^2/k$, and $a_2^2+b_2^2=-k(k_1+k_2)^2$. The above function is singular when the function $f(\theta)=e^{2\theta}-2\gamma e^{\theta}+1$ vanishes. It becomes zero when $e^{\theta}=\gamma\pm\sqrt{\gamma^2-1}$. Hence if $\gamma<1$ the solution is non-singular.

\section{Standard and Nonlocal MKdV Equations}

Letting $a_{2}=0$ and $a_{3}=i/a$ we get the mKdV system
\begin{eqnarray}
&& a q_{t}=\frac{1}{4}\, q_{xxx} -\frac{3}{2}\, q r q_{x}, \label{eq10} \\
&& a r_{t}=\frac{1}{4}\, r_{xxx} -\frac{3}{2}\, q r r_{x}. \label{eq11}
\end{eqnarray}
This system has the same recursion operator (\ref{recur}) as the NLS system. One-soliton solution of the above system is \cite{IWHI}
\begin{equation}\label{mKdVsystemonesol}
\displaystyle q(t,x)=\frac{e^{\theta_1}}{1+Ae^{\theta_1+\theta_2}}, \quad \quad r(t,x)=\frac{e^{\theta_2}}{1+Ae^{\theta_1+\theta_2}},
\end{equation}
with $\theta_i=k_ix-(k_i^3t/4a)+\delta_i$, $i=1, 2$, and $ A=-1/(k_1+k_2)^2$. Here $k_{1}$, $k_{2}$, $\delta_{1},$ and $\delta_{2}$ are arbitrary complex numbers. In mKdV case, there are also two types of approaches represented in \cite{GurPek2} to find solutions
of the standard mKdV and nonlocal mKdV (and cmKdV) equations.\\

\noindent 1. \,MKdV Equations: Let $r(t,x)=k q(t,x)$ then mKdV system reduces to the integrable mKdV equation
\begin{equation}
a q_{t}=\frac{1}{4}\, q_{xxx} -\frac{3 k}{2}\, q^2\, q_{x}. \label{eq12}
\end{equation}
In Type 1 one-soliton solution is obtained by letting $k_1=k_2=\alpha+i\beta$ and $e^{\delta_2}=ke^{\delta_1}=a_1+ib_1$ for
 $\alpha, \beta, a_1, b_1\in \mathbb{R}$ in (\ref{mKdVsystemonesol}) as
\begin{equation}\displaystyle
q(t,x)=\frac{e^{(\alpha+i\beta)x-\frac{(\alpha^3-3\alpha\beta^2)+i(3\alpha^2\beta-\beta^3)}{4a}t}(a_1+ib_1)}
{1-\frac{k}{4(\alpha^2+\beta^2)^2}e^{2(\alpha+i\beta)x-\frac{(\alpha^3-3\alpha\beta^2)+i(3\alpha^2\beta-\beta^3)}{2a}t}(a_1+ib_1)^2(\alpha-i\beta)^2}.
\end{equation}
Therefore we obtain the function
\begin{equation}\label{localone3}\displaystyle
|q(t,x)|^2=\frac{Y}{W},
\end{equation}
where
\begin{eqnarray}\displaystyle
Y&=&e^{2\alpha x-\frac{(\alpha^3-3\alpha\beta^2)}{2a}t}(a_1^2+b_1^2),\nonumber\\
W&=&1-\gamma_1\cos\theta+\frac{\gamma_1^2}{4}e^{\phi}=\frac{\gamma_1^2}{4}\Big[\frac{4}{\gamma_1^2}(1-\gamma_1\cos\theta)+e^{\phi}\Big],
\end{eqnarray}
where
$$\theta=2\beta x-\frac{1}{2a}(3\alpha^2\beta-\beta^3)t+\omega_0,\quad \phi=4\alpha x-\frac{1}{a}(\alpha^3-3\alpha\beta^2)t,$$
 for
$$\omega_0=\arccos(((a_1\alpha+b_1\beta)^2-(a_1\beta-b_1\alpha)^2)/(a_1^2+b_1^2)(\alpha^2+\beta^2))$$
and $\gamma_1=k(a_1^2+b_1^2)/2(\alpha^2+\beta^2)$. Hence we conclude that if $|\gamma_1|\leq 1$ the solution (\ref{localone3}) is non-singular. Type 2 approach gives $k_1=k_2=0$ yielding trivial solution.\\

\noindent 2.\, CmKdV Equations: Let $r=k \bar{q}(t,x)$ then mKdV system reduces to the integrable cmKdV equation
\begin{equation}
 a q_{t}=\frac{1}{4}\, q_{xxx} -\frac{3 k}{2}\, q\,\bar{q}\, q_{x}, \label{eq13}
\end{equation}
where $\bar{a}=a$. One-soliton solution is obtained by letting $k_2=\bar{k}_1=\alpha-i\beta$ for $\alpha, \beta \in \mathbb{R}$ and $e^{\delta_2}=ke^{\bar{\delta}_1}$ in (\ref{mKdVsystemonesol}) in Type 1 as
\begin{equation}\displaystyle
q(t,x)=\frac{e^{(\alpha+i\beta)x-\frac{(\alpha^3-3\alpha\beta^2)+i(3\alpha^2\beta-\beta^3)}{4a}t+\delta_1}}
{1-\frac{k}{4\alpha^2}e^{2\alpha x+\frac{(3\alpha\beta^2-\alpha^3)}{2a}t+\delta_1+\bar{\delta}_1}},
\end{equation}
so the function  $|q(t,x)|^2$ is
\begin{equation}\label{localone1}\displaystyle
|q(t,x)|^2=\frac{e^{2\alpha x-\frac{(\alpha^3-3\alpha\beta^2)}{2a}t+\delta_1+\bar{\delta}_1}}{(1-\frac{k}{4\alpha^2}e^{2\alpha x+\frac{(3\alpha\beta^2-\alpha^3)}{2a}t+\delta_1+\bar{\delta}_1})^2}.
\end{equation}
For $k< 0$, the solution (\ref{localone1}) can be written as
\begin{equation}\displaystyle
|q(t,x)|^2=-\frac{\alpha^2}{k}\mathrm{sech}^2\Big(\alpha x+\frac{(3\alpha \beta^2-\alpha^3)}{4a}t+\frac{\delta_1+\bar{\delta}_1}{2}+\delta\Big),
\end{equation}
where $\delta=\ln(-k/4\alpha^2)/2$. The above solution is non-singular.\\

\noindent We obtain a different one-soliton solution in Type 2 under the constraints $k_1=-\bar{k}_1$, $k_2=-\bar{k}_2$, $Ake^{\delta_1+\bar{\delta}_1}=1$, and $Ae^{\delta_2+\bar{\delta}_2}=k$ used in (\ref{mKdVsystemonesol}). If we let $k_1=\alpha i$, $k_2=\beta i$,
$e^{\delta_1}=a_1+ib_1$, and $e^{\delta_2}=a_2+ib_2$ for $\alpha, \beta, a_j, b_j\in \mathbb{R}$, $j=1, 2$, one-soliton solution
becomes
\begin{equation}\displaystyle
q(t,x)=\frac{e^{i\alpha x+i\frac{\alpha^3}{4a}t}(a_1+ib_1)}{1+\frac{1}{(\alpha+\beta)^2}e^{i(\alpha+\beta)x+i\frac{(\alpha^3+\beta^3)}{4a}t}(a_1+ib_1)(a_2+ib_2)},
\end{equation}
hence the corresponding function $|q(t,x)|^2$ is
\begin{equation}\displaystyle
|q(t,x)|^2=\frac{a_1^2+b_1^2}{4}\sec^2\Big(\frac{\theta}{2}\Big),
\end{equation}
where
$$\theta=(\alpha+\beta)x+\frac{1}{4a}(\alpha^3+\beta^3)t+\omega_0,$$
for $\omega_0=\arccos((a_1a_2-b_1b_2)/(\alpha+\beta)^2)$ with $a_1^2+b_1^2=(\alpha+\beta)^2/k$ and $a_2^2+b_2^2=k(\alpha+\beta)^2$. This is a singular solution for $\theta=(2n+1)\pi$, $n$ is an integer.\\

\noindent There are also two different types of nonlocal reductions.\\

\noindent 1. \,Nonlocal MKdV Equations: Let $r=k q (\mu_{1} t, \mu_{2} x)$ then mKdV system reduces to the integrable nonlocal mKdV equation
\begin{equation}
a q_{t}(t,x)=\frac{1}{4}\, q_{xxx}(t,x) -\frac{3 k}{2}\, q(t,x)\,q(\mu_{1} t, \mu_{2} x) q_{x}(t,x), \label{eq13}
\end{equation}
provided that $\mu_{1}\, \mu_{2}=1$. There is only one possibility $(\mu_{1}, \mu_{2})=(-1,-1)$. If we consider the Type 1 approach,
we get $k_1=-k_2$ which gives trivial solution $q(t,x)=0$. In Type 2, one-soliton solution
is obtained from (\ref{mKdVsystemonesol}) with the parameters satisfying the relations $Ake^{2\delta_1}=1$ and $Ae^{2\delta_2}=k$ as
\begin{equation}\label{realstsoln}\displaystyle
q(t,x)=\frac{i\sigma_1 e^{k_1x-\frac{k_1^3}{4a}t}(k_1+k_2)}{\sqrt{k}(1+\sigma_1\sigma_2e^{(k_1+k_2)x-\frac{(k_1^3+k_2^3)}{4a}t})}, \quad \sigma_j=\pm 1,\quad j=1, 2.
\end{equation}
If we let $a\in \mathbb{R}$, $k_1=\alpha_1+i\beta_1$, and $k_2=\alpha_2+i\beta_2$ then we obtain the solution $|q(t,x)|^2$ corresponding to (\ref{realstsoln}) as
\begin{equation}\label{realstsolnexp}
|q(t,x)|^2=\frac{e^{\theta}}{2k[\cosh(\phi)+\sigma_1\sigma_2\cos(\varphi)]},
\end{equation}
where $\theta=(\alpha_1-\alpha_2)x-((\alpha_1^3-3\alpha_1\beta_1^2-\alpha_2^3+3\alpha_2\beta_2^2)t/4a)$,
$\phi=A_1x+B_1t$, and $\varphi=A_2x+B_2t$. Here
\begin{eqnarray*}
&& A_1=\alpha_1+\alpha_2,\quad B_1=-\frac{1}{4a}(\alpha_1^3-3\alpha_1\beta_1^2+\alpha_2^3-3\alpha_2\beta_2^2),\\
&& A_2=\beta_1+\beta_2,\quad B_2=\frac{1}{4a}(\beta_1^3-3\alpha_1^2\beta_1+\beta_2^3-3\alpha_2^2\beta_2).
\end{eqnarray*}
\noindent There are cases where the solution (\ref{realstsolnexp}) is nonsingular:\\

\noindent
{\bf (a)} If we have $k_1=k_2$ for real $k_1$ and $\sigma_1\sigma_2=1$  then the solution (\ref{realstsoln}) becomes
\begin{equation}
q(t,x)=\frac{i\sigma_1k_1}{\sqrt{k}}\mathrm{sech}(k_1x-\frac{k_1^3}{4a}t).
\end{equation}

\noindent {\bf (b)} If $B_{1} A_{2}=B_{2} A_{1}$ then the solution (\ref{realstsolnexp}) becomes

\begin{equation}
|q(t,x)|^2=\frac{e^{\theta}}{2k[\cosh(\phi)+\sigma_1\sigma_2\cos(\frac{B_{2}}{B_{1}}\,\phi)]}.
\end{equation}

\begin{wrapfigure}{r}{0.3\textwidth}
  \vspace{-21pt}
  \begin{center}
    \includegraphics[width=0.18\textwidth]{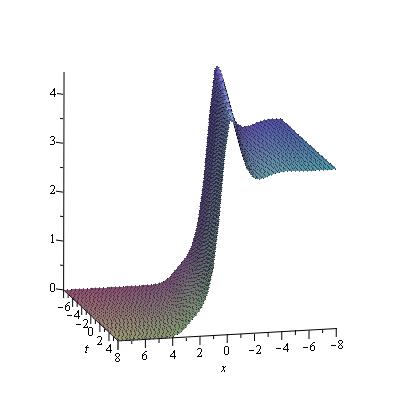}
  \end{center}
  \vspace{-18pt}
  \caption{A complexiton solution}
  \vspace{-10pt}
\end{wrapfigure}

\noindent{\bf Example 2.} If we take $(k_1, k_2,\sigma_1,\sigma_2 k, a)=(i, 1+(i/2),1, 1, -1, 1/4)$ then we have the solution
$$|q(t,x)|^2=\frac{13e^{-u}}{8[\cosh(u)+\cos(3u/2)]},$$
where $u=x-t/4$. This is a complexiton solution. The graph of this solution is given in Figure 2.\\

\vspace{0.5cm}

\noindent 2.\, Nonlocal CmKdV Equations: Let $r=k \bar{q}(\mu_{1} t, \mu_{2} x)$ then mKdV system reduces to the integrable nonlocal cmKdV equation
\begin{equation}
 a q_{t}(t,x)=\frac{1}{4}\, q_{xxx}(t,x) -\frac{3 k}{2}\, q(t,x)\,\bar{q}(\mu_{1} t, \mu_{2} x) q_{x}(t,x), \label{eq14}
\end{equation}
provided that $\bar{a}=\mu_{1}\, \mu_{2}\,a$. One-soliton solution is obtained by letting $k_2=\mu_2\bar{k}_1$ and $e^{\delta_2}=ke^{\bar{\delta}_1}$ in Type 1. In Type 2, a different one-soliton solution is obtained by letting $k_1=-\bar{k}_1\mu_2$, $k_2=-\bar{k}_2\mu_2$, $Ake^{\delta_1+\bar{\delta}_1}=1$, and $Ae^{\delta_2+\bar{\delta}_2}=k$. In this case there are three possibilities $(\mu_{1},\mu_{2})=\{(-1,1),(1,-1),(-1,-1)\}$.
Hence we have three integrable nonlocal cmKdV equations:\\

\noindent 2.i)\, T-Symmetric Nonlocal CmKdV Equations: Let $r=k \bar{q}(-t, x)$ then mKdV system reduces to the nonlocal cmKdV equation
\begin{equation}\displaystyle
aq_t(t,x)=-\frac{1}{4}q_{xxx}(t,x)+\frac{3}{2}k\bar{q}(-t,x)q(t,x)q_x(t,x), ~~~\bar{a}=-a.
\end{equation}
In Type 1 if we let $a=ib$, for nonzero $b\in \mathbb{R}$, $k_1=\alpha+i\beta$ so $k_2=\alpha-i\beta$ for $\alpha, \beta \in \mathbb{R}$, $\alpha\neq 0$ then one-soliton solution becomes
\begin{equation}\displaystyle
q(t,x)=\frac{e^{(\alpha+i\beta)x+\frac{i(\alpha^3-3\alpha\beta^2)-3\alpha^2\beta+\beta^3}{4b}t+\delta_1} }{1-\frac{k}{4\alpha^2}e^{2\alpha x+i\frac{\alpha^3-3\alpha\beta^2}{2b}t+\delta_1+\bar{\delta}_1 }   }.
\end{equation}
The corresponding function $|q(t,x)|^2$ is
\begin{equation}\label{nonlocalonegena}\displaystyle
|q(t,x)|^2=\frac{e^{2\alpha x+\frac{(\beta^3-3\alpha^2\beta)}{2b}t+\delta_1+\bar{\delta}_1 }}{\Big[\frac{k}{4\alpha^2}e^{2\alpha x+\delta_1+\bar{\delta}_1}
-\cos(\frac{(\alpha^3-3\alpha\beta^2)}{2b}t) \Big]^2+\sin^2(\frac{(\alpha^3-3\alpha\beta^2)}{2b}t) }.
\end{equation}
\noindent  When $\alpha^3-3\alpha\beta^2\neq 0$ and
$$t=\frac{2nb\pi}{(\alpha^3-3\alpha\beta^2)},\quad \frac{k}{4\alpha^2}e^{2\alpha x+\delta_1+\bar{\delta}_1}-(-1)^n=0,$$
where $n$ is an integer, for both focusing and defocusing cases,
the solution is singular. When $\alpha^3-3\alpha\beta^2=0$ the solution for focusing case is non-singular. When $\alpha=0$ the solution
is exponentially growing for $\beta^3/b>0$ and exponentially decaying for $\beta^3/b<0$.\\

\noindent In Type 2 if we let $a=i\alpha$, $k_1=i\beta$, $k_2=i\gamma$ for $\alpha, \beta,\gamma \in \mathbb{R}$, and $e^{\delta_1}=a_1+ib_1$, $e^{\delta_2}=a_2+ib_2$ for $a_j, b_j \in \mathbb{R}$, $j=1, 2$ then one-soliton solution becomes
\begin{equation}\displaystyle
q(t,x)=\frac{e^{i\beta x+\frac{\beta^3}{4\alpha}t}(a_1+ib_1)}{1+\frac{1}{(\beta+\gamma)^2}e^{i(\beta+\gamma)x+\frac{(\beta^3+\gamma^3)}{4\alpha}t}(a_1+ib_1)(a_2+ib_2)}.
\end{equation}
Hence the function $|q(t,x)|^2$ is
\begin{equation}\label{nonlocalonetype2a}\displaystyle
|q(t,x)|^2=\frac{e^{\frac{(\beta^3-\gamma^3)}{4\alpha}t}(a_1^2+b_1^2)}{2\Big[\cosh(\frac{(\beta^3+\gamma^3)}{4\alpha}t)+\cos\theta\Big]},
\end{equation}
where
$$\theta=(\beta+\gamma)x+\omega_0$$
for
$\omega_0=\arccos((a_1a_2-b_1b_2)/(\beta+\gamma)^2)$ with $a_1^2+b_1^2=(\beta+\gamma)^2/k$, $a_2^2+b_2^2=k(\beta+\gamma)^2$, and $\beta\neq -\gamma$. This solution is singular only at $t=0$, $\theta=(2n+1)\pi$ for $n$ integer.\\

\noindent 2.ii)\, S-Symmetric Nonlocal CmKdV Equations: Let $r=k \bar{q}(t, -x)$ then mKdV system reduces to the nonlocal cmKdV equation
\begin{equation}\displaystyle
aq_t(t,x)=-\frac{1}{4}q_{xxx}(t,x)+\frac{3}{2}k\bar{q}(t,-x)q(t,x)q_x(t,x),~~~\bar{a}=-a.
\end{equation}
If we consider Type 1 and let $a=ib$ for nonzero $b \in \mathbb{R}$, $k_1=\alpha+i\beta$ and so $k_2=-\alpha+i\beta$ for $\alpha, \beta \in \mathbb{R}$, $\beta\neq 0$ then one-soliton solution becomes
\begin{equation}\displaystyle
q(t,x)=\frac{e^{(\alpha+i\beta)x+\frac{i\alpha^3-3\alpha^2\beta-3i\alpha\beta^2+\beta^3}{4b}t+\delta_1} }{1+\frac{k}{4\beta^2}e^{2i\beta x+i\frac{\alpha^3-3\alpha\beta^2}{2b}t+\delta_1+\bar{\delta}_1 }   },
\end{equation}
and so  $|q(t,x)|^2$ is
\begin{equation}\label{nonlocalonegenb}\displaystyle
|q(t,x)|^2=\frac{e^{2\alpha x+\frac{(\beta^3-3\alpha^2\beta)}{2b}t+\delta_1+\bar{\delta}_1}}{\Big[\frac{k}{4\beta^2}e^{\frac{(\beta^3-3\alpha^2\beta)}{2b}t+\delta_1+\bar{\delta}_1}+
\cos(2\beta x)    \Big]^2+\sin^2(2\beta x)}.
\end{equation}
For $x=n\pi/(2\beta)$ and $ke^{(\beta^3-3\alpha^2\beta)t/2b+\delta_1+\bar{\delta}_1}/(4\beta^2)+(-1)^n=0$, where $n$
is an integer, the solution is unbounded but for $\beta^2=3\alpha^2$ and $ke^{\delta_1+\bar{\delta}_1}/(4\beta^2)+(-1)^n\neq 0$
we have a periodical solution. For $\alpha=0$, the solution (\ref{nonlocalonegenb}) becomes
\begin{equation}
|q(t,x)|^2=\frac{e^{\delta_1+\bar{\delta}_1}}{\gamma[\sigma_k\cosh(\frac{\beta^3}{2b}t+\ln(\frac{|\gamma|}{2}))+\cos(2\beta x)]},
\end{equation}
where $\gamma=ke^{\delta_1+\bar{\delta}_1}/(2\beta^2)$, $\sigma_k=1$ if $k>0$, and $\sigma_k=-1$ if $k<0$. This solution is non-singular
for $|\gamma|>2$, $\beta^3/b>0$ and $|\gamma|<2$, $\beta^3/b<0$ for any $t\geq 0$.\\

\noindent For Type 2 if we let $a=i\alpha$, $\alpha\in \mathbb{R}$, $e^{\delta_1}=a_1+ib_1$, and $e^{\delta_2}=a_2+ib_2$ for $a_j, b_j \in \mathbb{R}$, $j=1, 2$ we obtain the one-soliton solution as
\begin{equation}\displaystyle
q(t,x)=\frac{e^{k_1x+i\frac{k_1^3}{4\alpha}t}(a_1+ib_1)}{1-\frac{1}{(k_1+k_2)^2}e^{(k_1+k_2)x+i\frac{(k_1^3+k_2^3)}{4\alpha}t}(a_1+ib_1)(a_2+ib_2)}.
\end{equation}
Therefore the function $|q(t,x)|^2$ is
\begin{equation}\label{nonlocalonetype2b}\displaystyle
|q(t,x)|^2=\frac{e^{(k_1-k_2)x}(a_1^2+b_1^2)}{2[\cosh((k_1+k_2)x)+\cos\theta]},
\end{equation}
where
$$\theta=\frac{1}{4}(k_1^3+k_2^3)t-\omega_0$$
for $\omega_0=\arccos((b_1b_2-a_1a_2)/(k_1+k_2)^2)$ with $a_1^2+b_1^2=-\frac{(k_1+k_2)^2}{k}$ and $a_2^2+b_2^2=-k(k_1+k_2)^2$, $k_1\neq -k_2$. This solution has singularity at $x=0$, $\theta=(2n+1)\pi$ for $n$ integer.\\

\noindent 2.iii)\, ST-Symmetric Nonlocal CmKdV Equations: Let $r=k \bar{q}(-t, -x)$ then mKdV system reduces to the nonlocal cmKdV equation
\begin{equation}\displaystyle
aq_t(t,x)=-\frac{1}{4}q_{xxx}(t,x)+\frac{3}{2}k\bar{q}(-t,-x)q(t,x)q_x(t,x),~~~\bar{a}=a.
\end{equation}

\noindent In Type 1 if we let $k_1=\alpha+i\beta$ and so $k_2=-\alpha+i\beta$ for $\alpha, \beta \in \mathbb{R}$, $\beta\neq 0$ the one-soliton solution $q(t,x)$ becomes
\begin{equation}\label{casecnonlocal}\displaystyle
q(t,x)=\frac{e^{(\alpha+i\beta)x-\frac{\alpha^3+3\alpha^2i\beta-3\alpha\beta^2-i\beta^3}{4a}t+\delta_1}}
{1+\frac{k}{4\beta^2}e^{2i\beta x-i\frac{(6\alpha^2\beta-2\beta^3)}{4a}t+\delta_1+\bar{\delta}_1}}.
\end{equation}
Then we obtain the function $|q(t,x)|^2$ as
\begin{equation}\label{nonlocalonegenc}\displaystyle
|q(t,x)|^2=\frac{e^{\theta}}{\mu\Big[(\frac{1}{\mu}+\frac{\mu}{4})+\cos\phi\Big]},
\end{equation}
  where $$\theta=2\alpha x+\frac{1}{2a}(3\alpha\beta^2-\alpha^3)t+\delta_1+\bar{\delta}_1, \phi=2\beta x+\frac{1}{2a}(\beta^3-3\alpha^2\beta)t,$$
  and $\mu=ke^{\delta_1+\bar{\delta}_1}/(2\beta^2)$. This solution is non-singular for all $\mu$ except $\mu=\pm 2$.\\

\noindent  For Type 2,
  if we take $e^{\delta_1}=a_1+ib_1$ and $e^{\delta_2}=a_2+ib_2$ for $a_j, b_j \in \mathbb{R}$, $j=1, 2$ we obtain the one-soliton solution as
\begin{equation}\displaystyle
q(t,x)=\frac{e^{k_1x-\frac{k_1^3}{4a}t}(a_1+ib_1)}{1-\frac{1}{(k_1+k_2)^2}e^{(k_1+k_2)x-\frac{(k_1^3+k_2^3)}{4a}t}(a_1+ib_1)(a_2+ib_2)},
\end{equation}
hence the function $|q(t,x)|^2$ is
\begin{equation}\label{nonlocalonetype2c}\displaystyle
|q(t,x)|^2=\frac{e^{\phi}}{1-2\gamma e^{\theta}+e^{2\theta}},
\end{equation}
where
$$\theta=(k_1+k_2)x-\frac{1}{4a}(k_1^3+k_2^3)t,\, \phi=2k_1x-\frac{k_1^3}{2a}t,$$
$\gamma=(a_1a_2-b_1b_2)/(k_1+k_2)^2$ with $a_1^2+b_1^2=-(k_1+k_2)^2/k$, and $a_2^2+b_2^2=-k(k_1+k_2)^2$, $k_1\neq -k_2$. The above function has singularity when $e^{\theta}=\gamma\pm\sqrt{\gamma^2-1}$. Hence for $\gamma<1$ and $k_2>k_1$ the solution is non-singular and bounded.\\

\begin{wrapfigure}{r}{0.3\textwidth}
  \vspace{-21pt}
  \begin{center}
    \includegraphics[width=0.18\textwidth]{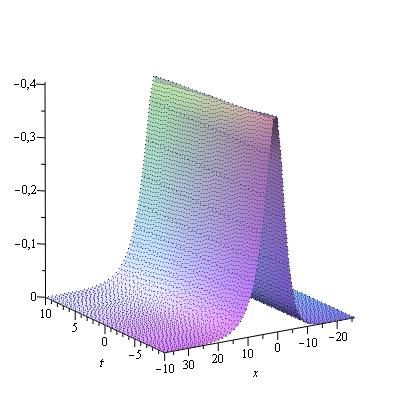}
  \end{center}
  \vspace{-18pt}
  \caption{An asymptotically decaying soliton}
  \vspace{-10pt}
\end{wrapfigure}

\noindent \textbf{Example 3.} For the set of the parameters $(k_1, k_2, e^{\delta_1}, e^{\delta_2},\\
 k, a)=(\frac{1}{2},\frac{1}{4}, -\frac{3}{4},\frac{3}{4}, -1, 2)$ we obtain the following asymptotically decaying soliton $$q(t,x)=\frac{(-3e^{\frac{1}{2}x-\frac{1}{64}t})}{4(1+e^{\frac{3}{4}x-\frac{9}{512}t})},$$
whose graph is given in Figure 3.\\

\noindent \textbf{Remark.} All dynamical variables considered so far are complex valued functions. We claim that all the results
presented here will be valid if the dynamical variables are pseudo complex valued functions. Any pseudo complex number is $\alpha=a+i b$ where $i^2=1$. Complex conjugation is  $\bar{\alpha}=a- i b$. Hence the norm of
a pseudo complex number is not positive definite $\alpha \bar{\alpha}=a^2-b^2$. NLS equation
\begin{equation}
i q_{t}=-\frac{1}{2} q_{xx}+k q^2 \bar{q},
\end{equation}
has real and imaginary parts  ($q=u+i v$)
\begin{eqnarray}
u_{t}=-\frac{1}{2} v_{xx}+k (u^2-\epsilon v^2) v, \nonumber\\
\epsilon v_{t}=-\frac{1}{2} u_{xx}+k (u^2-\epsilon v^2) u, \nonumber
\end{eqnarray}
where $i^2 =\epsilon=\pm 1$.

\section{Fordy-Kulish System}

Systems of integrable nonlinear partial differential equations arise when the Lax pairs are given in certain Lie algebras. Fordy-Kulish (FK) system of equations are examples of such equations \cite{fk}, \cite{for1}. We briefly give the Lax representations of these equations,
\begin{eqnarray}
\phi_{x}&=&(\lambda H_{S}+Q^{A}\, E_{A})\, \phi,  \label{lax1} \\
\phi_{t}&=&(A^a H_{a}+B^{A} E_{A}+C^{D} E_{D})\, \phi, \label{lax2}
\end{eqnarray}
where the dynamical variables are $Q_{A}=(q^{\alpha},p^{\alpha})$, the functions $A^{a}$, $B^{A}$, and $C^{A}$ depend on the spectral parameter
$\lambda$, on the dynamical variables ($q^{\alpha} $, $p^{\alpha}$), and their $x$-derivatives. The system of FK equations is an example when the functions $A$, $B$, and $C$ are quadratic functions of $\lambda$. Let $q^{\alpha}(t,x)$ and $p^{\alpha}(t,x)$ be the complex dynamical variables where $\alpha=1,2,\cdots, N$, then the FK integrable system arising from the integrability condition of Lax equations (\ref{lax1}) and (\ref{lax2})
is given by
\begin{eqnarray}
a q^{\alpha}_{t}&=& q^{\alpha}_{xx}+ R^{\alpha}\,_{\beta \gamma -\delta}\, q^{\beta}\,q^{\gamma}\, p^{\delta}, \label{denk5}\\
a p^{\alpha}_{t}&=& p^{\alpha}_{xx}+ R^{-\alpha}\,_{-\beta -\gamma \delta}\, p^{\beta}\,p^{\gamma}\, q^{\delta}, \label{denk6}
\end{eqnarray}
for all $\alpha=1,2,\cdots,N$. Here $R^{\alpha}\,_{\beta \gamma -\delta}$ and $R^{-\alpha}\,_{-\beta -\gamma \delta}$ are the curvature tensors of a Hermitian symmetric space satisfying
\begin{equation}
(R^{\alpha}\,_{\beta \gamma -\delta})^{\star}=R^{-\alpha}\,_{-\beta -\gamma \delta},\label{prop}
\end{equation}
and $a$ is a complex number. These equations are known as the FK system which is integrable in the sense that they are obtained from the zero curvature condition of a connection defined on a Hermitian symmetric space and these equations can also be written in a Hamiltonian form.\\

\noindent  The standard reduction of the above FK system is obtained by letting
$p^{\alpha}=k (q^{\alpha})^{\star}$
for all $\alpha=1,2,\cdots,N$. The FK system (\ref{denk5})-(\ref{denk6}) reduces to a single equation
\begin{equation}
a q^{\alpha}_{t}=q^{\alpha}_{xx} +k \, R^{\alpha}\,_{\beta \gamma -\delta}\, q^{\beta}\,q^{\gamma}\,(q^{\delta})^{\star}, ~~ \alpha=1,2,\cdots,N \label{denk7},
\end{equation}
provided that $a^{\star}=-a$ and (\ref{prop}) is satisfied. Here $*$ over a letter denotes complex conjugation.

\section{Nonlocal Fordy-Kulish Equations}

Here we will show that the Fordy-Kulish system is compatible with the nonlocal reduction of Ablowitz-Musslimani type. For this purpose
using a similar constraint as in NLS system we let
\begin{equation}
p^{\alpha}(t,x)=k [q^{\alpha}(\mu_{1} t, \mu_{2} x)]^{\star}, ~~~\alpha=1, 2, \cdots, N,
\end{equation}
where $\mu_{1}^2=\mu_{2}^2=1$. Under this constraint the FK system (\ref{denk5})-(\ref{denk6})
reduces to the following system of equations:
\begin{equation}
a q^{\alpha}_{t}(t,x)=q^{\alpha}_{xx}(t,x) +k \, R^{\alpha}\,_{\beta \gamma -\delta}\, q^{\beta}(t,x)\,q^{\gamma}(t,x)\,(q^{\delta}(\mu_{1} t, \mu_{2} x))^{\star},  \label{denk9}
\end{equation}
provided that $a^{\star}=-\mu_{1}\,a$ and (\ref{prop}) is satisfied. In addition to (\ref{denk9}) we have also an equation for $q^{\delta}(\mu_{1} t, \mu_{2} x)$ which can be obtained by letting $t \rightarrow \mu_{1}t$, $x \rightarrow \mu_{2}x$ in (\ref{denk9}). Hence we obtain T-symmetric, S-symmetric, and ST-symmetric nonlocal FK equations. Nonlocal reductions correspond to
$(\mu_{1}, \mu_{2})=\{(-1,1),(1,-1),(-1,-1)\}$. Hence corresponding to these values of $\mu_{1}$ and {$\mu_{2}$ we have three different nonlocal integrable FK equations. They are given as follows:\\

\noindent 1. T-Symmetric Nonlocal FK Equations:
\begin{equation}
a q^{\alpha}_{t}(t,x)=q^{\alpha}_{xx}(t,x)+ k \, R^{\alpha}\,_{\beta \gamma -\delta}\, q^{\beta}(t,x)\,q^{\gamma}(t,x)\,(q^{\delta}(-t,x))^{\star},  \label{denk10}
\end{equation}
with $a^{\star}=a$.\\

\noindent 2. S-Symmetric Nonlocal FK Equations:
\begin{equation}
a q^{\alpha}_{t}(t,x)=q^{\alpha}_{xx}(t,x)+ k \, R^{\alpha}\,_{\beta \gamma -\delta}\, q^{\beta}(t,x)\,q^{\gamma}(t,x)\,(q^{\delta}(t,-x))^{\star},  \label{denk11}
\end{equation}
with $a^{\star}=-a$.\\

\noindent 3. ST-Symmetric Nonlocal FK Equations:
\begin{equation}
a q^{\alpha}_{t}(t,x)=q^{\alpha}_{xx}(t,x) +k \, R^{\alpha}\,_{\beta \gamma -\delta}\, q^{\beta}(t,x)\,q^{\gamma}(t,x)\,(q^{\delta}(-t,-x))^{\star},  \label{denk11}
\end{equation}
with $a^{\star}=a$. All these three nonlocal equations are integrable.

\section{Super Integrable Systems}

When the Lax pair, in $(1+1)$-dimensions, is given in a super Lie algebra the resulting evolution equations are super integrable systems. They are given as
a coupled system
\begin{eqnarray}
q^{i}_{t}&=&F^{i}(q^{k}, \varepsilon^{k}, q^{k}_{x}, \varepsilon^{k}_{x}, q^{k}_{xx}, \varepsilon^{k}_{xx}, \cdots),~~~  \label{cdenk3}\\
 \nonumber \\
\varepsilon^{i}_{t}&=&G^{i}(q^{k}, \varepsilon^{k}, q^{k}_{x}, \varepsilon^{k}_{x}, q^{k}_{xx}, \varepsilon^{k}_{xx}, \cdots), \label{cdenk4}
\end{eqnarray}
for all $i=1,2,\cdots,N$ where $F^{i}$ and $G^{i} ~(i=1,2,\cdots,N)$ are functions of the dynamical variables $q^{i}(t,x)$, $\varepsilon^{i}(t,x)$, and their partial derivatives with respect to $x$. Here $q^{i}$'s are bosonic and $\varepsilon^{i}$'s are the fermionic dynamical variables. Since we start with a super Lax pair then the system (\ref{cdenk3})-(\ref{cdenk4}) is a super integrable system of nonlinear partial differential equations.

\section{Nonlocal Super NLS and MKdV Equations}

As an example taking the Lax pair in super $sl(2,R)$ algebra we obtain the super AKNS system. We have two bosonic $(q,r)$ and two fermionic ($\varepsilon, \beta$) dynamical variables. They satisfy the following evolution equations \cite{GO}-\cite{GOS},\\

\noindent Bosonic Equations:
\begin{eqnarray}
q_{t}&=&a_{2}\, (-\frac{1}{2}\, q_{xx}+q ^2\, r+2 \, \varepsilon_{x}\, \varepsilon+2 q \beta \varepsilon)+i a_{3}\,(-\frac{1}{4}\, q_{xxx}+\frac{3}{2}\, q r q_{x}+3 (\varepsilon_{x} \varepsilon)_{x}
\nonumber\\&&-3 q \beta_{x} \varepsilon +3 q \beta \varepsilon_{x}), \label{sup1} \\
 r_{t}&=&a_{2}\, (\frac{1}{2}\, r_{xx}-q \, r^2 +2\, \beta_{x}\, \beta-2 r \beta \varepsilon)+
i a_{3}\,(-\frac{1}{4}\, r_{xxx}+\frac{3}{2}\, q r r_{x}-3 (\beta_{x} \beta)_{x}\nonumber\\
 &&+3 r \beta_{x} \varepsilon -3 r \beta \varepsilon_{x}), \label{sup2}
\end{eqnarray}

\noindent Fermionic Equations:
\begin{eqnarray}
\beta_{t}&=&a_{2}\,(\beta_{xx}-r \varepsilon_{x}-\frac{1}{2}\, \varepsilon r_{x}-\frac{1}{2} q r \beta)+i a_{3} (-\beta_{xxx}+\frac{3}{4} r q_{x} \beta+\frac{3}{4} q r_{x} \beta+\frac{3}{2} q r \beta_{x}
\nonumber\\&&+\frac{3}{2} r_{x} \varepsilon_{x}+\frac{3}{4} \varepsilon r_{xx}), \label{sup3}\\
\varepsilon_{t}&=&a_{2}\,(-\varepsilon_{xx}+q \beta_{x}+\frac{1}{2}\, \beta q_{x}+\frac{1}{2} q r \varepsilon)
+i a_{3} (-\varepsilon_{xxx}+\frac{3}{4} r q_{x} \varepsilon+\frac{3}{4} q r_{x} \varepsilon +\frac{3}{2} q r \varepsilon_{x}
\nonumber\\&&+\frac{3}{2} q_{x} \beta_{x}+\frac{3}{4} \beta q_{xx}), \label{sup4}
\end{eqnarray}
where $a_{2}$ and $a_{3}$ are arbitrary constants.

\subsection{Super NLS Equations}

Letting $a_{3}=0$ in the equations (\ref{sup1})-(\ref{sup4}) we get the super coupled NLS system of equations. There are two bosonic ($q$,$r$) and two fermionic ($\varepsilon, \beta$) potentials satisfying
\begin{eqnarray}
a q_{t}&=&-\frac{1}{2}\,q_{xx}+q^2\,r+2 \varepsilon_{x}\, \varepsilon+2 q\, \beta\, \varepsilon, \label{seq1}\\
a r_{t}&=&\frac{1}{2}\,r_{xx}-q\,r^2+2 \beta_{x}\, \beta-2 r\, \beta\, \varepsilon, \label{seq2}\\
a \varepsilon_{t}&=&-\varepsilon_{xx}+q\, \beta_{x}+\frac{1}{2}\, \beta \, q_{x}+\frac{1}{2}\,q\,r \,\varepsilon, \label{seq3}\\
a \beta_{t}&=&\beta_{xx}-r\, \varepsilon_{x}-\frac{1}{2}\, \varepsilon \, r_{x}-\frac{1}{2}\,q\,r \,\beta, \label{seq4}
\end{eqnarray}
where $a_{2}=1/a$. The standard reduction is $r=k_{1}\, \bar{q}$ and $\beta=k_{2} \bar{\varepsilon}$ where  $k_{1}$ and $k_{2}$ are constants, a bar over a quantity denotes the Berezin conjugation in the Grassmann algebra. If $P$ and $Q$ are super functions (bosonic or fermionic) then
$\overline{{P}{Q}}={\overline{ Q}}\,{ \overline{P}}$. Under these constraints the above equations (\ref{seq1})-(\ref{seq4}) reduce to the following super NLS equations provided $k_{1}=k_{2}^2$ and $\bar{a}=-a$,
\begin{eqnarray}
a q_{t}&=&-\frac{1}{2}\,q_{xx}+k_{1}\, q^2\,\bar{q}+2 \varepsilon_{x}\, \varepsilon+2 k_{2}\,q\, \bar{\varepsilon}\, \varepsilon, \label{seq5}\\
a \varepsilon_{t}&=&-\varepsilon_{xx}+k_{2}\, q\, \bar{\varepsilon}_{x}+\frac{1}{2}\, k_{2}\, \bar{\varepsilon} \, q_{x}+ \frac{1}{2}\,k_{1}\,q\, \bar{q} \,\varepsilon. \label{seq6}
\end{eqnarray}
\noindent Here we show that super NLS system (\ref{seq1})-(\ref{seq4}) can be reduced to nonlocal super NLS equations. This can be done by choosing the super Ablowitz-Musslimani reduction as
\begin{equation}
 r(t,x)=k_{1}\, \bar{q}(\mu_{1} t, \mu_{2} x),~~~ \beta(t,x)= k_{2} \bar{\varepsilon}(\mu_{1} t, \mu_{2} x).
 \end{equation}
 where $\mu_{1}^2=\mu_{2}^2=1$. Here $k_{1}$ and $k_{2}$ are real constants. Under these constraints the above set (\ref{seq1})-(\ref{seq4}) reduces to super NLS equations \cite{kup}, \cite{kup1},
\begin{eqnarray}
a q_{t}(t,x)&=&-\frac{1}{2}\,q_{xx}(t,x)+k_{1}\,\, q^2(t,x)\,\bar{q}(\mu_{1} t, \mu_{2} x)+2 \varepsilon_{x}(t,x)\, \varepsilon(t,x)\nonumber\\
&&+2 k_{2}\,q(t,x)\, \bar{\varepsilon}(\mu_{1} t, \mu_{2} x) \varepsilon(t,x), \nonumber\\
a \varepsilon_{t}(t,x)&=&-\varepsilon_{xx}(t,x)+k_{2}\, q(t,x)\, \bar{\varepsilon}_{x}(\mu_{1} t, \mu_{2} x)+\frac{1}{2}\, k_{2} \, \bar{\varepsilon}(\mu_{1} t, \mu_{2} x) \, q_{x}(t,x)\nonumber\\
&&+ \frac{1}{2}\,k_{1}\,q(t,x)\, \bar{q}(\mu_{1} t, \mu_{2} x) \,\varepsilon(t,x),  \nonumber
\end{eqnarray}
provided that
\begin{equation}
\bar{a}\, \mu_{1}=-a,~~~ k_{2}^2\, \mu_{2}=k_{1}.
\end{equation}
Nonlocal reductions correspond to the choices $(\mu_{1}, \mu_{2})=\{(-1,1),(1,-1),(-1,-1)\}$. They are explicitly given by,\\

\noindent 1. T-Symmetric Nonlocal Super NLS Equations:
\begin{eqnarray*}
a q_{t}(t,x)&=&-\frac{1}{2}\,q_{xx}(t,x)+k_1\,\, q^2(t,x)\,\bar{q}(- t, x)+2 \varepsilon_{x}(t,x)\, \varepsilon(t,x)\nonumber\\
&&+2 k_2\,q(t,x)\, \bar{\varepsilon}(-t,x), \varepsilon(t,x), \label{seq13} \nonumber\\
a \varepsilon_{t}(t,x)&=&-\varepsilon_{xx}(t,x)+k_2\, q(t,x)\, \bar{\varepsilon}_{x}(-t,x)+\frac{1}{2}\, k_2 \, \bar{\varepsilon}(-t,x) \, q_{x}(t,x)
\nonumber\\
&&+ \frac{1}{2}\, k_1 \,q(t,x)\, \bar{q}(-t,x) \,\varepsilon(t,x),  \nonumber
\end{eqnarray*}
with $a^{\star}=a$ and $k_1=k_2^2$.\\

\noindent 2. S-Symmetric Nonlocal Super NLS Equations:
\begin{eqnarray*}
a q_{t}(t,x)&=&-\frac{1}{2}\,q_{xx}(t,x)+k_1\,\, q^2(t,x)\,\bar{q}(t, -x)+2 \varepsilon_{x}(t,x)\, \varepsilon(t,x)+2 k_2 \,q(t,x)\, \bar{\varepsilon}(t,-x) \varepsilon(t,x), \label{seq14}\\
a \varepsilon_{t}(t,x)&=&-\varepsilon_{xx}(t,x)+k_2\, q(t,x)\, \bar{\varepsilon}_{x}(t,-x)+\frac{1}{2}\, k_2 \, \bar{\varepsilon}(t,-x) \, q_{x}(t,x)
\nonumber\\
&&+ \frac{1}{2}\, k_1 \,q(t,x)\, \bar{q}(t,-x) \,\varepsilon(t,x),
\end{eqnarray*}
with $a^{\star}=-a$ and $k_1=-k_2^2$.\\

\noindent 3. ST-Symmetric Nonlocal Super NLS Equations:
\begin{eqnarray*}
a q_{t}(t,x)&=&-\frac{1}{2}\,q_{xx}(t,x)+k_1\,\, q^2(t,x)\,\bar{q}(-t, -x)+2 \varepsilon_{x}(t,x)\, \varepsilon(t,x)\\
&&+2 k_2\,q(t,x)\, \bar{\varepsilon}(-t,-x), \varepsilon(t,x), \label{seq15} \\
a \varepsilon_{t}(t,x)&=&-\varepsilon_{xx}(t,x)+k_2\, q(t,x)\, \bar{\varepsilon}_{x}(-t,-x)+\frac{1}{2}\, k_2 \, \bar{\varepsilon}(-t,-x) \, q_{x}(t,x)
\\&&+ \frac{1}{2}\,k_1 \,q(t,x)\, \bar{q}(-t,-x) \,\varepsilon(t,x),
\end{eqnarray*}
with $a^{\star}=a$ and $k_1=-k_2^2$.

\subsection{Super MKdV Systems}

Another special case of the super AKNS equations is the super mKdV system \cite{GO}, \cite{GO2}
\begin{eqnarray}
a q_{t}&=&-\frac{1}{4}\,q_{xxx}+\frac{3}{2}\,r\,q\,q_{x}+3 (\varepsilon_{x}\, \varepsilon)_{x}-3\, q\, \beta_{x}\, \varepsilon+3 q \beta\, \varepsilon_{x}, \nonumber\\
a r_{t}&=&-\frac{1}{4}\,r_{xxx}+\frac{3}{2}\,r\,q\,r_{x}-3 (\beta_{x}\, \beta)_{x}+3\, r\, \beta_{x}\, \varepsilon-3 r \beta\, \varepsilon_{x}, \label{seq10},\nonumber\\
a \varepsilon_{t}&=&-\varepsilon_{xxx}+ \frac{3}{4}\, (r\, q)_{x}\, \varepsilon+\frac{3}{2}\,q\,r\, \varepsilon_{x}+\frac{3}{2}\,q_{x}\, \,\beta_{x}+\frac{3}{4} \beta\, q_{xx}, \label{seq11} \nonumber\\
a \beta_{t}&=&-\beta_{xxx}+ \frac{3}{4} (r\, q)_{x}\, \beta+\frac{3}{2}\,q\,r\, \beta_{x}+\frac{3}{2}\,r_{x}\, \,\varepsilon_{x}+\frac{3}{4} \varepsilon\, r_{xx}. \nonumber
\end{eqnarray}
The standard reduction is $r=k_{1} \bar{q}, \beta=k_{2}\, \bar{\varepsilon}$. Then we obtain \cite{GO},
\begin{eqnarray}
a q_{t}&=&-\frac{1}{4}\,q_{xxx}+\frac{3}{2}\,k_{1}\, \bar{q}\,q\,q_{x}+3 (\varepsilon_{x}\, \varepsilon)_{x}-3\,k_{2} q\, \bar{\varepsilon}_{x}\, \varepsilon+3 k_{2}\,q \bar{\varepsilon}\, \varepsilon_{x}, \nonumber\\
a \varepsilon_{t}&=&-\varepsilon_{xxx}+ \frac{3}{4}\,k_{1} (\bar{q}\, q)_{x}\, \varepsilon+\frac{3}{2}\,k_{1}q\, \bar{q}\, \varepsilon_{x}+\frac{3}{2}\,k_{2}\, q_{x}\, \,\bar{\varepsilon}_{x}+\frac{3}{4}\,k_{2} \bar{\varepsilon}\, q_{xx}, \nonumber
\end{eqnarray}
provided that $k_{1}=k_{2}^2$ and $\bar{a}=a$. For the super mKdV system, Ablowitz-Musslimani type of reduction is also possible. Letting
\begin{equation}
 r(t,x)=k_1\, \bar{q}(\mu_{1} t, \mu_{2} x),~~~ \beta(t,x)= k_2 \bar{\varepsilon}(\mu_{1} t, \mu_{2} x),
 \end{equation}
 where $\mu_{1}^2=\mu_{2}^2=1$ we get the following system of equations
\begin{eqnarray}
a q_{t}(t,x)&=&-\frac{1}{4}\,q_{xxx}(t,x)+\frac{3}{2}\,k_1\, \bar{q}(\mu_{1} t, \mu_{2} x)\,q(t,x)\,q_{x}(t,x)
+3 (\varepsilon_{x}(t,x)\, \varepsilon(t,x))_{x} \nonumber \\
&&-3\, q(t,x)\, \bar{\varepsilon}_{x}(\mu_{1} t, \mu_{2} x)\, \varepsilon(t,x)+3 k_2\,q(t,x)\, \bar{\varepsilon}(\mu_{1} t, \mu_{2} x)\, \varepsilon_{x}(t,x), \label{seq17}\\
a \varepsilon_{t}(t,x)&=&-\varepsilon_{xxx}(t,x)+ \frac{3}{4}\,k_1 (\bar{q}(\mu_{1} t, \mu_{2} x)\, q(t,x))_{x}\, \varepsilon(t,x)+\frac{3}{2}\, k_1 q(t,x)\, \bar{q}(\mu_{1} t, \mu_{2} x)\, \varepsilon_{x}(t,x) \nonumber\\
&&+\frac{3}{2}\, k_2 \, q_{x}(t,x)\, \,\bar{\varepsilon}_{x}(\mu_{1} t, \mu_{2} x)
+\frac{3}{4}\,k_2 \bar{\varepsilon}(\mu_{1} t, \mu_{2} x)\, q_{xx}(t,x), \label{seq18}
\end{eqnarray}
provided that $\bar{a}\, \mu_{1}\, \mu_{2}=a$, $k_2^2 \mu_{2}=k_1$. Nonlocal reductions correspond to the choices $(\mu_{1}, \mu_{2})=\{(-1,1),(1,-1),(-1,-1)\}$. They are explicitly given by,\\

\noindent 1. T-Symmetric Nonlocal Super MKdV Equations: Here $\bar{a}=-a$ and $k_1=k_2^2$.
\begin{eqnarray}
a q_{t}(t,x)&=&-\frac{1}{4}\,q_{xxx}(t,x)+\frac{3}{2}\,k_1\, \bar{q}(-t,x)\,q(t,x)\,q_{x}(t,x)
+3 (\varepsilon_{x}(t,x)\, \varepsilon(t,x))_{x} \nonumber \\
&&-3\, q(t,x)\, \bar{\varepsilon}_{x}(-t,x)\, \varepsilon(t,x)+3 k_2\,q(t,x) \bar{\varepsilon}(-t,x)\, \varepsilon_{x}(t,x), \label{seq119}\\
a \varepsilon_{t}(t,x)&=&-\varepsilon_{xxx}(t,x)+ \frac{3}{4}\,k_1 (\bar{q}(-t,x)\, q(t,x))_{x}\, \varepsilon(t,x)+\frac{3}{2}\,k_1 q(t,x)\, \bar{q}(-t,x)\, \varepsilon_{x}(t,x) \nonumber\\
&&+\frac{3}{2}\,k_2 \, q_{x}(t,x)\, \,\bar{\varepsilon}_{x}(-t,x)
+\frac{3}{4}\,k_2 \bar{\varepsilon}(-t,x)\, q_{xx}(t,x), \label{seq20}
\end{eqnarray}

\noindent 2. S-Symmetric Nonlocal Super MKdV Equations: Here $\bar{a}=-a$ and $k_1=-k_2^2$.
\begin{eqnarray}
a q_{t}(t,x)&=&-\frac{1}{4}\,q_{xxx}(t,x)+\frac{3}{2}\, k_1\, \bar{q}(t,-x)\,q(t,x)\,q_{x}(t,x)
+3 (\varepsilon_{x}(t,x)\, \varepsilon(t,x))_{x} \nonumber \\
&&-3\, q(t,x)\, \bar{\varepsilon}_{x}(t,-x)\, \varepsilon(t,x)+3 k_2\,q(t,x) \bar{\varepsilon}(t,-x)\, \varepsilon_{x}(t,x), \label{seq119}\\
a \varepsilon_{t}(t,x)&=&-\varepsilon_{xxx}(t,x)+ \frac{3}{4}\,k_1 (\bar{q}(t,-x)\, q(t,x))_{x}\, \varepsilon(t,x)+\frac{3}{2}\, k_1 q(t,x)\, \bar{q}(t,-x)\, \varepsilon_{x}(t,x) \nonumber\\
&&+\frac{3}{2}\, k_2 \, q_{x}(t,x)\, \,\bar{\varepsilon}_{x}(t,-x)
+\frac{3}{4}\, k_2 \bar{\varepsilon}(t,-x)\, q_{xx}(t,x), \label{seq20}
\end{eqnarray}

\noindent 3. ST-Symmetric Nonlocal Super MKdV Equations: Here $\bar{a}=a$ and $k_1=-k_2^2$.
\begin{eqnarray}
a q_{t}(t,x)&=&-\frac{1}{4}\,q_{xxx}(t,x)+\frac{3}{2}\,k_1\, \bar{q}(-t,-x)\,q(t,x)\,q_{x}(t,x)
+3 (\varepsilon_{x}(t,x)\, \varepsilon(t,x))_{x} \nonumber\\
&&-3\, q(t,x)\, \bar{\varepsilon}_{x}(-t,-x)\, \varepsilon(t,x)+3 k_2\,q(t,x) \bar{\varepsilon}(-t,-x)\, \varepsilon_{x}(t,x), \label{seq119}\\
a \varepsilon_{t}(t,x)&=&-\varepsilon_{xxx}(t,x)+ \frac{3}{4}\, k_1 (\bar{q}(-t,-x)\, q(t,x))_{x}\, \varepsilon(t,x)+\frac{3}{2}\, k_1 q(t,x)\, \bar{q}(-t,-x)\, \varepsilon_{x}(t,x) \nonumber\\
&&+\frac{3}{2}\, k_2 \, q_{x}(t,x)\, \,\bar{\varepsilon}_{x}(-t,-x)
+\frac{3}{4}\, k_2 \bar{\varepsilon}(-t,-x)\, q_{xx}(t,x), \label{seq20}
\end{eqnarray}

\section{Concluding Remarks}

In this work we first presented all integrable nonlocal reductions of NLS and MKdV systems. We gave the recursion operators and the soliton solutions
of these nonlocal equations. We then presented the extension of the nonlocal NLS equation to nonlocal Fordy-Kulish equations on symmetric
spaces. Starting with the super AKNS system we studied all
possible nonlocal reductions and found two new super integrable systems. They are the nonlocal super
NLS equations and nonlocal super mKdV systems of equations. There are three different nonlocal
types of super integrable equations. They correspond to T-, S-, and ST- symmetric super
NLS and super mKdV equations.

From the study of NLS and mKdV systems (both bosonic and fermionic integrable systems) we observed that they  have standard and nonlocal reductions. Moreover in both of these systems there are at least one nonlocal reduction to a standard reduction. For instance both systems have $r(t,x)=k \bar{q}(t,x)$ as a standard reduction and the corresponding nonlocal reductions are $r(t,x)=k \bar{q}(\mu_{1} t,\mu_{2}x)$ where $k$ is real constant and $ (\mu_{1}, \mu_{2})=(1,-1),(-1,1), (-1,-1)$. From these reductions we obtain standard and nonlocal NLS equations and standard and nonlocal complex mKdV equations and their nonlocal super integrable extensions. The mKdV system has additional standard and nonlocal reductions. Standard reduction $r(t,x)=k q(t,x)$, $k$ is real constant, and its corresponding nonlocal reduction $r(t,x)=k q(-t,-x)$ give the nonlocal mKdV equation. From all these experiences we conclude with a conjecture: If a system of equations admits a standard reduction then there exists at least one corresponding nonlocal reduction of the same system.

\section{Acknowledgment}
 This work is partially supported by the Scientific
and Technological Research Council of Turkey (T\"{U}B\.{I}TAK).

\end{document}